\documentclass[sigconf]{acmart}

\usepackage{booktabs} 
\copyrightyear{2018}
\acmYear{2018}
\setcopyright{rightsretained}
\acmConference[SIGIR '18]{The 41st International ACM SIGIR Conference
on Research & Development in Information Retrieval}{July 8--12,
2018}{Ann Arbor, MI, USA}
\acmBooktitle{SIGIR '18: The 41st International ACM SIGIR Conference on
Research \& Development in Information Retrieval, July 8--12, 2018, Ann
Arbor, MI, USA}
\acmDOI{10.1145/3209978.3210195}
\acmISBN{978-1-4503-5657-2/18/07}

\begin{document}

\title{DATA:SEARCH'18 -- Searching Data on the Web}

\author{Paul Groth}
\affiliation{%
  \institution{Elsevier Labs}
  \city{Amsterdam}
  \country{The Netherlands}
  \postcode{43017-6221}
}
\email{p.groth@elsevier.com}

\author{Laura Koesten}
\affiliation{%
  \institution{The Open Data Institute; University of Southampton}
  \streetaddress{P.O. Box 1212}
  \city{London}
  \country{United Kingdom}
  \postcode{43017-6221}
}
\email{laura.koesten@theodi.org}

\author{Philipp Mayr}
\affiliation{%
  \institution{GESIS -- Leibniz Institute for the Social Sciences}
  \city{Cologne}
  \country{Germany}
}
\email{philipp.mayr@gesis.org}

\author{Maarten de Rijke}
\orcid{0000-0002-1086-0202}
\affiliation{%
  \institution{University of Amsterdam}
  \streetaddress{P.O. Box 1212}
  \city{Amsterdam}
  \country{The Netherlands}
  \postcode{}
}
\email{derijke@uva.nl}

\author{Elena Simperl}
\affiliation{%
  \institution{University of Southampton}
  \streetaddress{P.O. Box 1212}
  \city{Southampton}
  \country{United Kingdom}
  \postcode{43017-6221}
}
\email{e.simperl@soton.ac.uk}

\renewcommand{\shortauthors}{P. Groth et al.}

\begin{abstract}
This half day workshop explores challenges in data search, with a particular focus on data on the web. We want to stimulate an interdisciplinary discussion around how to improve the description, discovery, ranking and presentation of structured and semi-structured data, across data formats and domain applications. We welcome contributions describing algorithms and systems, as well as frameworks and studies in human data interaction. The workshop aims to bring together communities interested in making the web of data more discoverable, easier to search and more user friendly.
\end{abstract}

\keywords{Data search, dataset retrieval, human data interaction, web of data}

\maketitle

\section{Organising committee}
\begin{itemize}
\item Paul Groth, Elsevier Labs, The Netherlands
\item Laura Koesten, The Open Data Institute, United Kingdom
\item Philipp Mayr, GESIS -- Leibniz Institute for the Social Sciences, Germany
\item Maarten de Rijke, University of Amsterdam, The Netherlands
\item Elena Simperl, University of Southampton, United Kingdom
\end{itemize}

\section{Motivation and relevance}

\subsection{Background}
As an increasing amount of data becomes available on the web, searching for it becomes an increasingly important, timely topic~\citep{gregory-2018-understanding}. The web hosts a whole range of new data species, published in structured and semi-structured formats - from web markup using schema.org and web tables to open government data portals, knowledge bases such as Wikidata and scientific data repositories~\cite{IDC:report,Lehmberg2016}. This data fuels many novel applications, for example fact checkers and question answering systems, and enables advances in machine learning and AI. 

Just like any other resources on the web, data benefits from network effects - it becomes more useful, and creates more value, when it is discoverable. And yet, despite advances in information retrieval, the Semantic Web and data management, data search is by far not as advanced, both technologically ~\citep{Cafarella:2011:SDW:1897816.1897839} and from a user experience point of view ~\citep{Koesten2017}, as related areas such as document search.

Most approaches to user-centric data search are domain-specific or have been created with certain task contexts, data schemas or data formats in mind~\citep{dai-2017-learning}. 
Conducting research to explore dataset search outside these constraints is both important and timely for a venue such as SIGIR. 
The aim of the workshop is to be a venue to present and exchange ideas and experiences for discovering and searching all types of structured or semi-structured datasets and to discuss how concepts and lessons learned from academic search, entity search, digital libraries, and web search could be transferred to data search scenarios. 

The opportunities to share and establish links between different perspectives on search and discovery for different kinds of data are significant and can inform the design of a wide range of information retrieval technologies, including search engines, recommender systems and conversational agents. 

A broad range of methods and insights are important to enable the discovery of, and access to, data published on the web, including
\begin{itemize}
\item analyzing contextual information for datasets, including mentions of datasets
\item browsing and query support for structured and semi-structured data
\item inference and data enrichment systems
\item learning to match for datasets
\item learning to rank datasets
\item mining direct links between documents, datasets or data records
\item summaries and descriptions of datasets targeting users or search engines
\item concepts and methods to present data and entity-centric results.
\end{itemize}

We see a large space for discussion and future research in the development of federated data discovery and search technologies, which leverages the most recent advances in information retrieval, Semantic Web and databases, and is mindful of human factors.

\subsection{Recent developments}
Dataset search and discovery has emerged in a range of complementary disciplines.
\citet{Kunze2013} introduce dataset retrieval as a specialization of information retrieval, however, they restrict their scope to the process of returning relevant RDF datasets. 

Several architectures have been proposed to support discoverability of web data. The Linked Data community has put forward a set of principles and technologies to publish data in a form that makes it easy for applications to find and reuse it. Publishers are encouraged to define links between datasets, which can be used, alongside de-referencing URIs, to find additional data. The Linked Open Data Cloud, as well as large knowledge graphs such as Wikidata and DBpedia are prime examples of this approach. 

A related development are data portals, for example in open government and some scientific domains, and data sharing networks, such as Kaggle and data.world. Data portals are centralized repositories where an institution, or  multiple institutions, provide access to their datasets. They classify the datasets according to pre-defined categories, support basic keyword and faceted search features, and present dataset results via short descriptions, metadata and sometimes visualizations. In a search log analysis of open data portals, \citet{Kacprzak2017} found that queries issued on data portals differ from those issued to web search engines in their length and structure. Data sharing networks use a social network paradigm to help people discover new datasets and engage with data publishers and users. 

Dataset search might be construed as just another type of entity search, like expert finding~\citep{balog-expertise-2012} or product search~\citep{vangysel-learning-2016}. 
However, \citet{DBLP:conf/adcs/ThomasOR15} show that dataset repositories have poor search over and inside tables. It is difficult for a user to tell from a repository's portal whether a useful dataset is available, and this problem is only likely to get worse. \citeauthor{DBLP:conf/adcs/ThomasOR15} demonstrate that the na\"{\i}ve approach of full-text search is not appropriate. They describe an alternative, based on inferring types of data and indexing columns as a unit, and demonstrate some improvements in early success especially when long captions are not available.
New retrieval models are needed, models, moreover, that can be optimized with limited training and/or interaction data~\cite{dai-2017-learning,carevic_JCDL2018}.

Data requires context to create meaning ~\citep{dervin1997given}, to \emph{make sense of it}. This is dependent on people's data literacy, technical skills and prior knowledge.~\citet{kelly2009methods} also shows that individuals vary significantly in terms of cognitive makeup, prior knowledge and behavioral dispositions. While this applies to search for all information sources generally, literature suggests unique characteristics when the information source is structured data.
In user studies with social scientists,~\citet{kern2015} found that the quantity and quality of metadata are far more critical in dataset search than in literature search, where convenience is most important. For empirical social scientists, the choice of research data was found to be more relevant than the choice of literature; therefore they were  willing to put more effort into the retrieval process.
In a mixed methods study describing the information seeking process for structured data, \citet{Koesten2017} combined in-depth interviews with data professionals and a search log analysis of a large open governmental data portal. They note that finding data is challenging, even for data professionals who are familiar with state of the art tools, and that data search is often exploratory and complex. Evaluation criteria for datasets in a search scenario show unique characteristics -- the importance of context alongside information about provenance and methods for collection and analysis emerged as key factors, which help professionals determine whether a dataset is relevant and useful for their purposes \citep{Koesten2017,GregoryGCSW17}.

\section{Theme and purpose}
The objective of this workshop was to bring together researchers and practitioners interested in advancing data search on the web. This includes looking at the specifics of data-centric information seeking behavior, understanding interaction challenges in data search on the web, and analyzing the cognitive processes involved in the consumption of structured data by users. At the same time, we aimed to discuss architectures and technologies for data search - including semantics and information retrieval for structured and semi-structured data (e.g., ranking algorithms and indexing), in particular in the context of decentralized and distributed systems such as the web. We are interested in approaches to analyze, characterize and discover data sources. We want to facilitate a continuing discussion around data search across formats and domain-specific applications.

We envisioned the workshop as a forum for researchers and practitioners from various disciplines to come together and discuss common challenges and identify synergies for joint initiatives.

\subsection{Topics}
DATA:SEARCH'18\footnote{\url{https://datasearch-ws.github.io/2018/}} sought application-oriented papers, as well as more theoretical papers, position papers and empirical studies. 

The workshop proposed a multidisciplinary discussion on the following themes, with a focus on search and discovery of RDF, CSV, JSON and other structured and semi-structured data sources:

\begin{itemize}
\item Analyzing behavioral traces during data search
\item Approaches to personalization and contextualization in data\-set search
\item Data summarization
\item Dataset representation for retrieval (standards, models, work\-arounds)
\item Decentralized and distributed architectures and algorithms in data search
\item Deep linking of datasets
\item Entity recognition in datasets
\item Evaluation of dataset search tools and algorithms
\item Fusing, cleaning, ranking and refining dataset search results
\item Information seeking behavior for data (interactive data retrieval)
\item Data indexing and profiling approaches
\item Learning to rank for data search
\item Query routing taking into account relevance, quality and profiles of distributed datasets
\item Retrieval models for data search
\item Scalability and performance of distributed data queries
\item Search results presentation for datasets
\item Visual and speech interfaces to datasets
\item Semantic dataset search
\item Usability of data portals and data discovery tools
\item User modeling for data search
\item Systems and user studies in data search in vertical domains, including transport, geospatial data, science, weather etc.
\end{itemize}

We encouraged contributions using a variety of methods. This can include, for example, user studies, lab experiments, system-based evaluations, but also experiments using gamification and crowdsourcing.

The workshop was organized around a keynote, a set of lightening talks and round table discussions.

\section{Organizers}

Datasets as information objects are situated at the intersection of several disciplines -- information retrieval, semantic web, user interaction, and library science. 
Discovery of, and access to datasets, is an emerging shared interest of academic and industrial researchers.
The team behind this workshop proposal represents all of these angles and interests, both in the organizers and in the proposed program committee.

\subsection{Co-chairs}
\noindent\textbf{Dr Paul Groth} is Disruptive Technology Director at Elsevier Labs. He has done research at the University of Southern California and the Vrije Universiteit Amsterdam. His research focuses on intelligent systems for dealing with large amounts of diverse contextualized knowledge with a particular focus on web and science applications. This includes research in data provenance, data science, data integration and knowledge sharing. Paul is co-author of "Provenance: an Introduction to PROV" as well as over a 100 peer-reviewed publications. He has chaired multiple international events including Beyond the PDF 2, The International Semantic Web Conference, and the International Provenance and Annotation Workshop.\\
More info: http://pgroth.com. Elsevier Labs; email: p.groth@elsevier.\-com\\

\noindent\textbf{Laura Koesten} is a Maria Curie Sk\l odowska fellow, doing her PhD at the Open Data Institute and at the University of Southampton in the UK. She is part of WDAqua, a European Union's Horizon 2020 initiative to advance state of the art Question Answering. Her research interests are Human Computer Interaction, Interactive Information Retrieval with a focus on dataset retrieval, Open Data and Semantic Interfaces. In her PhD she is looking at ways to improve Human Data Interaction in IIR systems. She publishes at CHI and has a background in Human Factors, with an MSc degree from Loughborough University.\\
The Open Data Institute; University of Southampton, UK;  Email: laura.koesten@theodi.org\\

\noindent \textbf{Dr Philipp Mayr}
is a deputy department head and a team leader at the GESIS -- Leibniz Institute for the Social Sciences department Knowledge Technologies for the Social Sciences (WTS). He received his PhD in applied informetrics and information retrieval from the Berlin School of Library and Information Science at Humboldt University Berlin in 2009. To date, he has been awarded substantial research funding (PI, Co-PI) from national and European funding agencies. Philipp has published in top conferences and prestigious journals in the areas informetrics, information retrieval and digital libraries. His research interests include: interactive information retrieval, scholarly recommendation systems, data retrieval, non-textual ranking, bibliometric and scientometric methods, applied informetrics, science models in digital libraries, knowledge representation, semantic technologies, user studies, information behavior. 
More info: \url{https://philippmayr.github.io/}.
GESIS -- Leibniz Institute for the Social Sciences, Email: philipp.mayr@gesis.org\\

\noindent\textbf{Prof Maarten de Rijke}
is professor of Computer Science at the University of Amsterdam. He is a member of the Royal Dutch Academy of Arts and Sciences (KNAW), a former director of Amsterdam Data Science, a collaborative network involving 600+ data scientists in the Amsterdam area, and the founding director of ICAI, the Innovation Center for Artificial Intelligence. 
His research is situated at the interface of AI and IR, and focused on learning to rank, semantic search, and autonomous environments for information interaction.\\
More info: \url{https://staff.fnwi.uva.nl/m.derijke/}.
University of Amsterdam. Email: derijke@uva.nl\\

\noindent\textbf{Prof Elena Simperl} is professor of Computer Science at the University of Southampton. Her research interests include knowledge engineering, Social Web technologies, and crowdsourcing. She has contributed and led over 20 national and European research projects and authored more than 100 scientific publications, and chaired the European Semantic Web Conference (ESWC) in 2011 and 2012 and International Semantic Web Conference (ISWC) in 2016. She was vice-president of STI International until 2016 and the director of the ESWC summer school series. She has co-chaired more than 15 workshops, including the series on Theory and Practice of Social Machines (SOCM) at WWW, Crowdsourcing the Semantic Web at ISWC and Ontology Engineering in a Data Driven World at EKAW.\\
More info: \url{http://elenasimperl.eu/}.
University of Southampton, UK; Email: e.simperl@soton.ac.uk

\subsection{Programme Committee}

A list of PC members is as follows:
\begin{itemize}
\item Alexander Kotov (Wayne State University)
\item Arjen de Vries (Radboud University Nijmegen)
\item Arno Scharl (Modul University Vienna)
\item Axel Polleres (Vienna University of Economics and Business)
\item Eva M\'{e}ndez (Open research data)
\item Kuansan Wang (Microsoft)
\item Laura Dietz (University of New Hampshire)
\item Michael Gubanov (University of Texas, San Antonio)
\item Peter Haase (Metaphacts)
\item Steffen Lohmann (Fraunhofer IAIS)\\
\end{itemize}

\section{Related workshops}
Some of the organizers will be involved in the delivery of a workshop on data profiling and search in April 2018 at the Web Conference.\footnote{\url{https://profiles-datasearch.github.io/2018/}} That workshop is a follow-up of a workshop that has been run for many years at the European and International Semantic Web Conference, which focused on semantic techniques to enrich datasets to help with tasks such as discovery, description and sense-making of entity-centric data on the web.\footnote{\url{http://data4urbanmobility.l3s.uni-hannover.de/index.php/en/2017/11/07/profiles-workshop-iswc-2017/}} 

While we see many synergies between DATA:SEARCH'18 and those events, the focus of DATA:SEARCH'18 is on search. We aim to gain a better understanding of the extent to which techniques, methods and lessons learned from document retrieval broadly construed could apply to data-centric contexts, and explore in more depth the differences between the two areas from a technical and interaction perspective. 

\begin{acks}
This research was partially supported by the European Union's Horizon $2020$ research and innovation programme (under the Marie Sk\l{}odowska-Curie grant ID 642795) and They Buy For You (grant ID 780247),
EPSRC (Datastories, grant ID EP/P025676/1),
Ahold Delhaize,
Amsterdam Data Science,
the Bloomberg Research Grant program,
the China Scholarship Council,
the Criteo Faculty Research Award program,
DFG, grant no. SU 647/19-1, the OSCOSS project,
Elsevier,
the European Community's Seventh Framework Programme (FP7/2007-2013) under
grant agreement nr 312827 (VOX-Pol),
the Google Faculty Research Awards program,
the Microsoft Research Ph.D.\ program,
the Netherlands Institute for Sound and Vision,
the Netherlands Organisation for Scientific Research (NWO)
under pro\-ject nrs
CI-14-25, 
652.\-002.\-001, 
612.\-001.\-551, 
652.\-001.\-003, 
and
Yandex.

All content represents the opinion of the authors, which is not necessarily shared or endorsed by their respective employers and/or sponsors.
\end{acks}

\bibliographystyle{ACM-Reference-Format}
\bibliography{sample-bibliography}

\end{document}